\documentclass[letterpaper,draft]{aipprocx}
\layoutstyle{6sx}

\begin{document}

\title[A Brief History of the SCP]{The Acceleration of the Expansion of the Universe: A Brief Early History of the Supernova Cosmology Project (SCP)} 

\author{Gerson Goldhaber\footnote{gerson@lbl.gov}}{address={Lawrence Berkeley National Laboratory and Physics Department University of California, Berkeley, CA 94720}}


\begin{abstract}
It is now about 10 years since the evidence, based on Type Ia supernovae,  for the acceleration of the expansion of the Universe was discovered. I will discuss some aspects of the work and events in the Supernova Cosmology Project (SCP), during   the period 1988 to 1998, which led to this discovery.
\end{abstract}

\maketitle

Talk presented at Dark Matter Conference DM08, Marina del Rey, Feb. 20, 2008.

\section{Preamble}

 In the past I have written papers that described new physics results.  In this report, however, on this tenth anniversary of the discovery of dark energy, I present the history of our discovery \emph{as I saw it}.  I will recount parts of this many-year project that I remember particularly vividly as important in the work I was doing.   Of course, the other members of the team would recall different mixes of 
events and  efforts in which they were particularly engaged, and I hope they will find the opportunity to give their recollections.

When the bubble chamber work in Luis Alvarez's group at LBNL came to an end in the early 1970s, there started an interest in astrophysics. Rich Muller became interested in the Cosmic Microwave Background (CMB). He was later joined by George Smoot. After convincing Luis Alvarez of the feasibility of a CMB measurement and obtaining his support,  Rich Muller and George Smoot and  their collaborators studied the CMB with detectors placed on U2 airplane flights. This led to their discovery of the CMB dipole asymmetry. Smoot then went on to work with the COBE satellite, and was the leader in  the discovery of the CMB anisotropy. 

Meanwhile, Rich Muller, joined by Carl Pennypacker (a staff member at the Space Science Laboratory), set up an automated search for nearby supernovae (SNe). This work was primarily under Pennypacker's responsibility and was helped by Richard Treffers of the Berkeley Astronomy department and others. They were  later  joined by Muller's student Saul Perlmutter who carried out his thesis, a search for "Nemesis", the suspected companion star to the sun, as well as developing the
analyses and software for the SNe search.  During the period 1980-1988, they demonstrated that the method, originally suggested by Stirling Colgate et al. \cite{colgate75a}, worked, and they were able to discover about 25 nearby SNe. This work became the prototype of later automated SNe searches.

In 1988, a National Science Foundation (NSF) Center for Particle  Astrophysics  (CfPA) was being formed on the UC Berkeley campus.  Pennypacker  and  Perlmutter, working in Muller's group, had developed a new experiment to discover the ``Fate of the Universe'' through a study of distant Type Ia SNe, claimed to be ``standard candles.'' This experiment was included as one of the elements of the new CfPA \cite{pennypacker88a}.

\section{The SCP Group and the Discovery of Distant SNe}

In 1989, as  I was thinking about  my next experiment,\footnote{In 1989 my co-group leader George Trilling and I had come to an end of a nearly twenty-year collaboration with Burton Richter and Martin Perl and coworkers  of SLAC and LBL where we had discovered the psion family, charmed mesons, and the tau lepton among many others.} I was invited  by Carl Pennypacker to join in the search for the discovery of the ``Fate of the Universe'' in what was at that time called the "Deep Supernova Search."  The subject appealed to me, as did the proposed technique, since it involved evaluating images, something I have been doing throughout my career. \footnote{ Evaluating images is something I have been doing throughout my physics career, beginning with photographic emulsions in the 1950s, to bubble chambers in the 1960s,
to computer-reconstructed particle events in the 1970s, and 1980s.  Thus, the supernova experiment with its computer-reconstructed optical images, seemed 
to be a nice fit with my experience and inclinations.} 

\begin{figure}[ht]
\resizebox{0.8\textwidth}{!}{\includegraphics{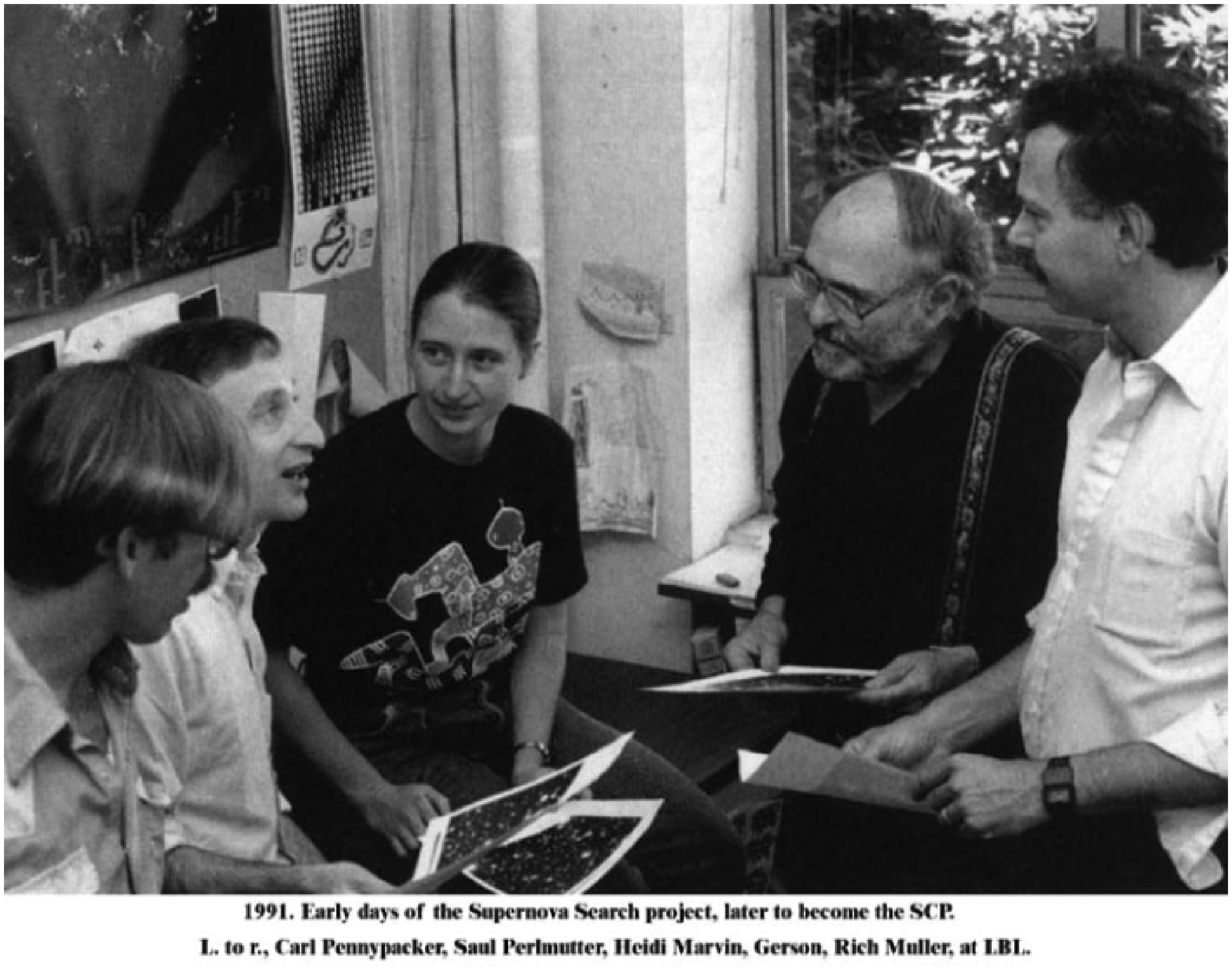}}
\caption{The SCP group at Berkeley in 1990-91. Carl Pennypacker, Saul Perlmutter, Heidi Marvin, Gerson Goldhaber, and Rich Muller l. to  r. }
\label{fig:1}
\end{figure}

In 1990-91 the entire group at Berkeley (now known as the Supernova Cosmology Project (SCP)), consisted of Carl Pennypacker, Saul Perlmutter, Heidi Marvin, myself, and Rich Muller (shown in Fig.~\ref{fig:1}, l. to r.) . 

Soon after my joining the group, there were major changes in its makeup. In 1991 Rich Muller decided to spend his efforts in research on the ice ages and global weather patterns, and Carl Pennypacker founded the "Hands on Universe" Program for high school students and began to devote most of his time to educational activities. This left Saul and me to carry on with graduate student Heidi Marvin Newberg and later Alex Kim.  Shortly thereafter, Bob Cahn, at the time the physics division director at LBNL, discussed with me the question of a group leader. With my strong support, he decided to appoint Saul to that position.

 Perhaps because Rich, Saul, Carl, and I were not established as astronomers, it was at first difficult for us to get time on the large premier telescopes. Before I joined the group, on Carl's initiative , Carl, Rich and Saul, together with two colleagues in Australia, Brian Boyle (starting out in the UK) and Warrick Couch, did manage to obtain scheduled time on the 3.9m Anglo Australian Telescope and built a focus reducing lens and CCD camera for it. During the three years we observed at this telescope, while the system worked well (and became one of the most-used instruments at the telescope for years after),  there was no identified  SN candidate.  Unfortunately, 80\% of our scheduled nights were lost due to bad weather. The lens and camera construction and installation and the techniques we developed, as well as an unidentified  candidate found later at the 2.5m Isaac Newton Telescope, are discussed in great detail in Heidi Marvin Newberg's thesis \cite{newberg89a}. 
 
 While for the first three years of our efforts we did not find any identified supernovae,\footnote{However, our general approach received important validation with the
discovery by Norgaard-Nielsen et
al. \cite{norgaard-nielsen89a} of a single high z supernova (red shift of 0.31) after a two year effort. This SN was however discovered well after maximum light.} this time was not wasted, as it allowed us to develop the techniques which led to the eventual success of the project.

\begin{figure}[!htbp]
\resizebox{1.1\textwidth}{!}{\includegraphics{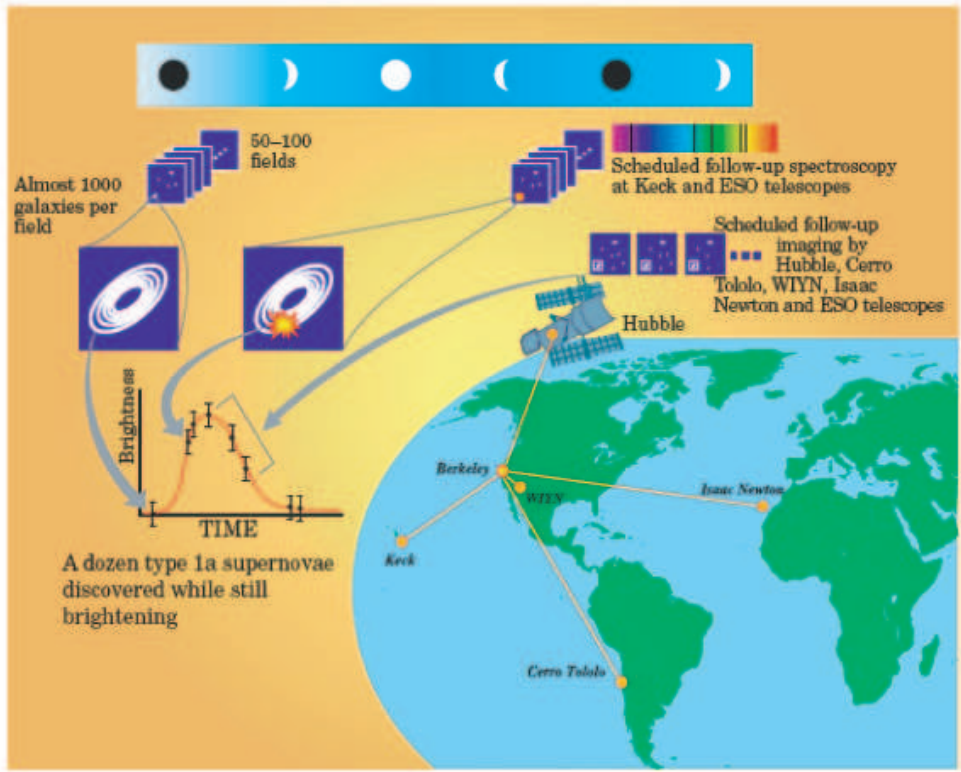}}
\caption{The strategy developed by Saul Perlmutter for finding SNe on demand from repeated CCD images.}
\label{fig:2}
\end{figure}

In particular, Saul Perlmutter developed the technique for finding ``SNe on demand'' or ``batch mode". This involved taking a  CCD ``reference" image just after a new moon, and coming back to take CCD ``discovery'' images before the next new moon (See Fig.~\ref{fig:2}). The data were sent back to Berkeley the same night they were  taken at the telescope. On the next morning we ran a program that subtracted the reference image from the discovery image and flagged possible candidates. In a few hours of hand scanning we were able to select the promising SN candidates. Thus as soon as SN candidates were discovered one could send observer(s) to the Keck 10m telescope to measure spectra and, after that, start to take data points for the light curve at ground based telescopes. (See Fig.~\ref{fig:2} for list of follow-up telescopes.)  For the first time, new supernovae could be guaranteed to be discovered on a certain date, and while they were still brightening--making it possible to propose for scheduled nights on the follow-up telescopes. This technique, or some variation of it,  was adopted by all subsequent  SN  searches.

It is instructive to compare Saul's new technique to what was being invented as the state of the art in the early 1990s at much lower redshifts, Hamuy et al.'s Calan/Tololo Supernova Search (CTSS).  This search observed fields (with photographic plates or film) twice a month, and then organized follow-up observation campaigns -- but unlike Saul's technique is was not aimed at producing guaranteed batches of (on-the-rise) SN discoveries on a given date.   Thus, towards the end of the CTSS search, Hamuy et al. \cite{hamuy93b}    wrote:  ``Unfortunately, the appearance of a SN is not predictable. As a consequence of this we cannot schedule the followup observations {\em a priori}, and we generally have to rely on someone else's telescope time. This makes the execution of this project somewhat difficult.''

    As pointed out in a review article by Perlmutter and Schmidt, Saul's technique addressed a different problem: ``The SCP targeted a much higher redshift range, $z > 0.3$, in order to measure the (presumed) deceleration of the Universe, so it faced a different challenge than the CTSS search. The high redshift SNe required discovery, spectroscopic confirmation, and photometric follow up on much larger telescopes. This precious telescope time could neither be borrowed from other visiting observers and staff nor applied for in sufficient quantities spread throughout the year to cover all SNe discovered in a given search field, and with observations early enough to establish their peak brightness. Moreover, since the observing time to confirm high redshift SNe was significant on the largest telescopes, there was a clear `chicken and egg' problem: telescope time assignment committees would not award follow-up time for a SN discovery that might, or might not, happen on a given run (and might, or might not, be well past maximum) and, without the follow-up time, it was impossible to demonstrate that high redshift SNe were being discovered by the SCP''   (Perlmutter \& Schmidt \cite{perlmutter03a}). Saul's technique solved this problem.

After we found the first few SNe,  Saul was able to convince the various telescope scheduling committees to give us time at the matching epochs to carry out this approach. In particular this not only allowed us to get more time to do searching, but also to get multi-band photometric and
spectroscopic follow-up.  By 1996 we were even allotted time at the Hubble Space Telescope, where, beginning in 1997, we measured lightcurve data points for some of the most reliably identified SNe. 

To eliminate cosmic rays and hot pixels, we took two images, usually five minutes apart, both for reference images and discovery images. Early in 1990, while scanning for SNe candidates (but   not finding any!) I did discover a group of about 20 asteroids.  The asteroids are distinguished by the fact that they are seen to move between the two discovery images.  However, some of them do not move significantly, and can be confused with SNe.  We therefore introduced a change in our procedures, interspersing a different field between the two discovery images, so as to allow an additional interval of 5 to 10 minutes between them. This allowed easy identification of the asteroids. Asteroids were later used by Carl Pennypacker in his Hands on Universe program to allow students to discover and measure real astronomical objects.  

In 1991  we were joined by visiting scholars Ariel Goobar and Silvia Gabi
 from Sweden and CERN and later on by Reynald Pain from France and Isobel Hook, an astronomer with expertise in spectroscopy, from England. 

In 1993 we collaborated with Alexei Filippenko, of the UC Berkeley Astronomy Department,  for measurements of spectra at the  Keck Telescope in Hawaii. 

In 1994 Don Groom, from  the Particle Data Group at LBL, joined our group,  as well as Susana Deustua.  Don Groom adapted the CERN program MINUIT to fit  our SNe  to the light curve and to determine the stretch parameter. 

Graduate students Alex Kim,  Matthew Kim and programmer Ivan Small were also essential in this early work. 

By 1995, now that we had became successful in finding distant supernovae, we were able to hire two postdocs, Rob Knop and Peter Nugent, and later, Greg Aldering.  These new people, as well as Susana, all had a background in astronomy.

By 1998 our group had grown to an international collaboration with 32 members\footnote{ Gregory    Aldering,
Brian    Boyle,
Patricia    Castro,
Warrick    Couch,
Susana    Deustua,
Richard    Ellis,
Sebastien    Fabbro,
Alexei    Filippenko,
Andrew    Fruchter,
Gerson    Goldhaber,
Ariel    Goobar,
Donald    Groom,
Isobel    Hook,
Mike    Irwin,
Alex    Kim,
Matthew    Kim,
Robert    Knop,
Julia    Lee,
Chris    Lidman,
Thomas    Matheson,
Richard    McMahon,
Heidi    Newberg,
Peter    Nugent,
Nelson    Nunes,
Reynald    Pain,
Nino    Panagia,
Carl    Pennypacker,
Saul    Perlmutter,
Robert    Quimby,
Pilar    Ruiz-Lapuente,
Brad    Schaefer,
Nicholas    Walton. 
(Alexei Filippenko moved to the Hi-Z Team when it was formed in 1995.)
} that signed the Dark Energy discovery paper \cite{perlmutter99a}. In 2007 Saul and the other 31  were co-recipients of the Gruber prize in cosmology for the discovery of Dark Energy, together with Brian Schmidt and the Hi-Z Team \cite{riess98a}.

\section{Milestones Leading to the Discovery of Dark Energy}

 Here is a list of some of the important comments and papers which allowed us to make our discovery. 
     
\begin{itemize}

\item  Suggestions to use Type Ia SNe as standard candles: From 1930s on Zwicky, Baade, Sandage, Kowal, Tammann etc.
\item 1984 Pskovskii \cite{pskovskii84a}, 1993 Phillips \cite{phillips93a}, 1995,1997 Perlmutter et al. \cite{perlmutter95b,ruiz-lapente97a,perlmutter97a}, 1995,2001 Goldhaber et al. \cite{perlmutter95b,ruiz-lapente97a,goldhaber01a} : light curve shape related to brightness and introduction of ``stretch''.   (An alternative approach to this light curve shape correction was in 1995, 1996 Riess, Press, and Kirshner \cite{riess95a,riess96a}. This approach, which employed a prior assumption on dust extinction, was used by the Hi-Z team.)
\item 1988 Leibundgut \cite{leibundgut88a}: Type Ia Lightcurve Template.
\item 1989 Norgaard-Nielsen et al. \cite{norgaard-nielsen89a}: First distant SN, z = 0.31 however discovered well past maximum light (maximum brightness).
\item  1990 Leibundgut \cite{leibundgut90a}, 1993 Hamuy et al. \cite{hamuy93a}: Single filter K-corrections.  To convert a filter region of the redshifted spectrum to  the same filter at zero redshift.
\item  1992 After three years of searching we discovered and followed our first distant SN (z = 0.458) before maximum light \cite{perlmutter95a}.
\item  1993-6 Hamuy, Phillips, Suntzeff, Maza et al. \cite{hamuy95a,hamuy96a}: CALAN/TOLOLO $z < 0.1$ reference sample of 29 Type Ia SNe. Of these we used 18 they discovered before 10 days past maximum light.
\item  1994-1995 Conference presentations by Perlmutter showing the effectiveness of the "SNe on demand" technique.
\item  1995 Kim, Goobar, and Perlmutter \cite{kim96a},   2002 Nugent, Kim, and Perlmutter \cite{nugent02a} :  Cross filter K-corrections.  By roughly matching the high-redshift observations' filter to the redshifted low-redshift observations' filter this approach solved the problem of the large statistical and systematic uncertainites in single-filter K corrections due to the large extrapolation errors when the exact form of the spectrum is not known and could not be practically measured over the full lightcurve. 
\item 1995 Goobar and Perlmutter \cite{goobar95a}: Separation of $\Omega_m$  and $\Omega_\Lambda$ shown to be feasible, fitting apparent magnitudes of SNe Ia at a variety of redshifts extending to $z > 1$.   This paper gives the luminosity distance formula for the Hubble plot calculations and shows that at each redshift the best-fit confidence region will make a diagonal on the $\Omega_m$ vs. $\Omega_\Lambda$ plane (with a slope that varies with redshift). 
\end{itemize}
Here "stretch", introduced by Saul, is the deviation of the measured timescale of a lightcurve from a standard average lightcurve. Typical values are $s= 0.8$ to $ 1.2 $. The larger $s$ the brighter the SN. The effective B-magnitudes are corrected to a standard lightcurve.

\section{Sept. 24, 1997: A Peak in the  $\Omega_m$ Histogram}

What has become the conventional way to determine $\Omega_m$ and  $\Omega_\Lambda$ is the Goobar-Perlmutter \cite{goobar95a}  best fit confidence level distribution on the  $\Omega_m$,  $\Omega_\Lambda$   plane.
This approach, favored by astronomers, requires accurate determinations  of the errors and correlated errors. When the SNe light curves are  first fitted  to a template this information is not yet available. 

The alternative approach, similar to the technique  used to find resonances in particle physics, is to look for peaks   in the distributions (on histograms)  of the variable being studied, namely mass in particle physics and $\Omega_m$ in our case here. In this approach the  measurement errors are reflected in the width of the resulting peak.
 
In September 1997 we had completed a first pass of the light curve data point analysis on 38 SNe, but the detailed error analyses were not yet available.
It took several months from the time a SN was discovered until the light curve points could be evaluated. We also took some final reference points a year after discovery. Rob Knop, who was working on this, was able typically to complete the measurements on one to two SNe per month.

  On the basis of these 38 Type Ia SNe we  obtained an indication that the universe is accelerating, rather than decelerating as we had originally expected. This was the culmination of nine years of work learning how to find high redshift SNe, how to measure them, how to fit them, how to K-correct them, how to stretch correct them, how to account for dust reddening, and then finally  fit them to a lightcurve template. This was all achieved by a collaborative effort of the entire group.  I studied the SNe after the data points on the lightcurve had been measured, fitted them with the group's lightcurve fitter which gave us the stretch value, and  made a table with some 20 attributes for each SN. 

\begin{figure}[!htbp]
\resizebox{0.8\textwidth}{!}{\includegraphics[angle=90]{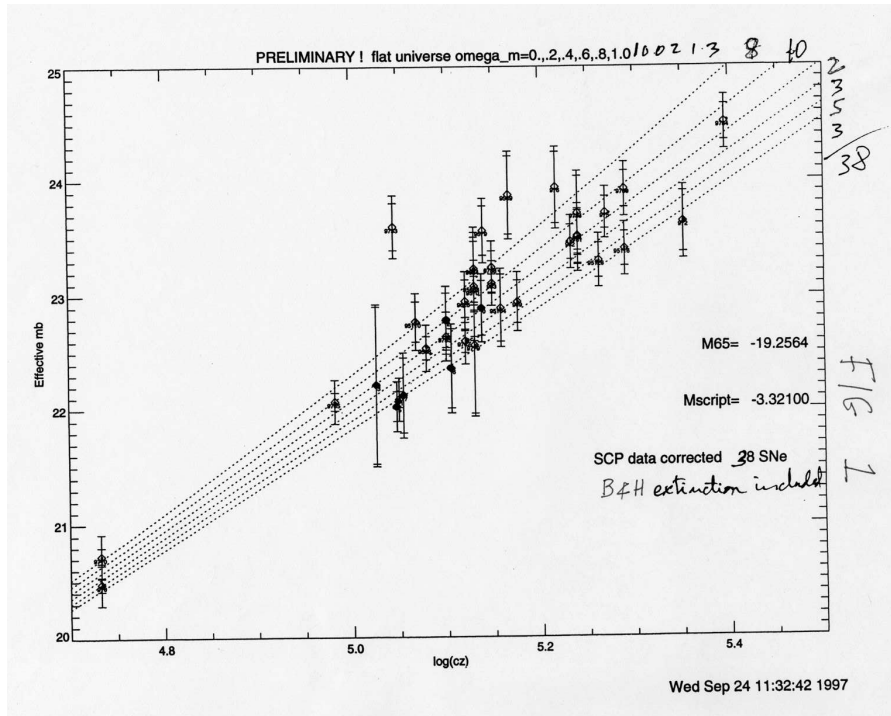}}
\caption{Our preliminary data of 38 SNe, as of Sep., 24, 1997. The effective B-magnitude vs. log(cz). Here the effective B-magnitude is the K-corrected and stretch-corrected magnitude and z is the redshift. The curves for a series of $\Omega_m$ values, in a flat universe, are shown. The sum of the data points in each $\Delta\Omega_m$ interval, as  counted by hand, is also shown at the right hand upper edge of the figure.}
\label{fig:3}
\end{figure}

\begin{figure} [!h]
\resizebox{0.8\textwidth}{!}{\includegraphics{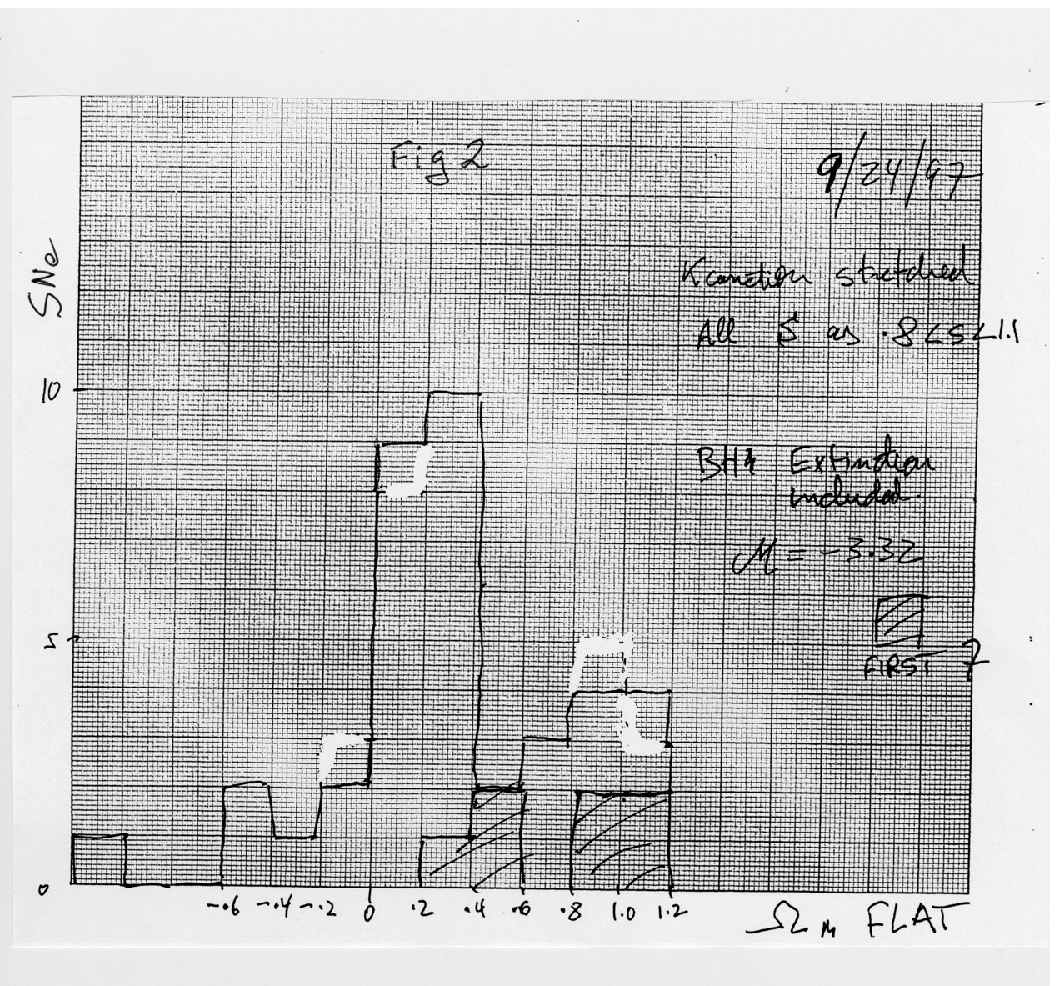}}
\caption{ The $\Omega_m$ distribution for the data points in Fig.~\ref{fig:3}, as presented at our SCP group meeting on Sep. 24, 1997,  is shown. The values for the first 7 SNe, which gave a considerably larger value for $\Omega_m$,  are shown cross-hatched. Here  $\Omega_\Lambda$ is given by $1- \Omega_m$.}
\label{fig:4}
\end{figure}

The question was next how to obtain the  $\Omega_m$ distribution from this plot. What I tried was to plot Hubble curves, for a flat universe, with  $\Omega_m$  in fine
intervals at 0.0, 0.2, 0.4, 0.6, 0.8, 1.0. See Fig.~\ref{fig:3}. A distribution in $\Omega_m$  is then obtained by counting (by hand!)  all the SNe that fell into each $\Delta\Omega_m$ interval giving the  histogram shown in Fig.~\ref{fig:4}. A few days later Saul wrote a program, SNLOOK, to have the computer count the number of SNe in each  $\Delta\Omega_m$ interval.  He also made plots both for a flat universe, and for a universe with $\Omega_\Lambda = 0$.  
 
Here it should be noted that a  flat universe was  not completely established at that point in time (1997). (The flat universe case was  based in part on inflation theory \cite{guth81a,linde82a,albrecht82a} and in part on very early rumors from  as yet unpublished CMB fluctuation data \cite{torbet99a}. The detailed measurements of the fluctuations in the CMB,  demonstrating a flat universe, came later \cite{hanany00a,debernardis00a,bennett03a}.) 
However, an analysis of the supernova data for the cases of a flat universe and a $\Omega_\Lambda = 0$ universe is a very informative way to characterize the data and its implications.  As the 1995 Goobar and Perlmutter paper\cite{goobar95a} had shown , the measurements of high-redshift supernovae in the redshift range we were studying result in a long diagonally-oriented confidence region in the $\Omega_m - \Omega_\Lambda$ plot, like the one shown in Figure 9.    We can also analyze this data by projecting onto an $\Omega_m$ histogram for either of the simple cosmology cases, the flat universe line or the $\Omega_\Lambda = 0$ line in Figure 9.  Given the  $\sim$45-degree slope of the diagonal confidence region for the SN redshift ranges we were studying, if we find a very low value of $\Omega_m$ in the flat-universe-case histogram, we will be guaranteed to find a negative value for $\Omega_m$ in the $\Omega_\Lambda=0$-case histogram.   Since there is only positive mass in our physics, this implies that there is a cosmological constant or some other form of Dark Energy -- and the result holds whether or not the universe is actually flat!   This is very different from the low $\Omega_m$ measurements from galaxy clusters, etc., that were being made in the late 1990s, where the measurement by itself did not tell you anything about $\Omega_\Lambda$.

In Fig.~\ref{fig:4} the peak we actually saw at   $\Omega_m = 0.2$ to $0.3$ for a flat universe indicates that $\Omega_\Lambda$ is   $0.8$ to $0.7$. Thus  we obtained a significant positive value for $\Omega_\Lambda$ or possibly a cosmological constant. As we were looking for deceleration we always calculated the deceleration parameter $q_0 = (1/2)\Omega_m - \Omega_\Lambda$. Here  $q_0 = -0.7$ to $-0.55$. A negative value for $q_0$ implies acceleration. We calculated $q_0$ for each of our samples and it always said: ACCELERATION.

The low redshift distribution of SN absolute magnitudes available at that time already showed a strong peak that indicated a large fraction
of supernovae were not suffering extinction. The strong peak in the $\Omega_m$ distribution was likely
also to be due to unextincted SNe, but we understood that, whatever this extinction distribution, as long as
it was the same at low and high redshift the cosmology results would be correct. We studied
the color distributions at both redshift ranges, and found that the distributions were sufficiently
consistent. This histogram of Figure~\ref{fig:4} (and the others shown below ) therefore accounts for
dust in this statistical approach, rather than correcting extinction SN by SN.
(Our other analyses at
that time examined this point with further studies of the measured colors of the distant
SNe, performing analyses correcting individual SN brightnesses and analyses comparing
groups of SNe. We concluded that our cosmology result was robust -- and that the correct
way to analyze color and dust for all the data sets at that time was not to use a Baysean prior
on the dust distribution, as others had done.) It is also interesting to note that the stretch correction,
which becomes very important in later precision measurements, does not change these
preliminary values significantly. It only tends to broaden the observed peak. At any rate
we did apply the stretch correction to these data.

This startling result was naturally treated with some skepticism within the group, when these plots were discussed in the weekly SCP meetings at the end of September 1997.   Although we had already submitted a Nature paper\cite{perlmutter98a} indicating evidence for a cosmological constant, we were still surprised that this result was settling in on the lower end of the $\Omega_m$ range from our first seven SNe \cite{perlmutter95b,ruiz-lapente97a,perlmutter97a} which had centered on a high value for $\Omega_m$.\footnote{ In retrospect, it is instructive to revisit the question of what happened with those first seven supernovae. As shown shaded in Fig.~\ref{fig:4} they are all at high $\Omega_m$ values.  Two turned out to be outliers with one probably not a Type Ia.  The other 5  moved slightly towards lower mass densities with remeasurement, but for the most part they simply happened to lie on the tail of the distribution. The lesson is:  beware of statistics of small numbers!}  These  seven SNe  are shown as cross hatched in Fig.~\ref{fig:4}. and indeed occur at the upper end of the   distribution.

 Again the fact that the deceleration coefficient was negative, indicating acceleration, was hard to swallow at first. I had been ``bump hunting'' for the past 30 years, and found the observed peak completely convincing. I believe that Saul convinced himself  after he wrote the  program,  SNLOOK and  by independently studying  our data. He later in October and December presented  such a histogram in colloquia  giving    $\Omega_m$ and $\Omega_\Lambda$. 
 Some other colleagues, Rob Knop and  Carl Pennypacker,  appeared convinced; while Greg Aldering was not fully convinced, he stated however that this histogram ``helped to galvanize the effort within our group". Indeed, as the  result of a major effort, the SCP group was able to calculate the best fit  confidence levels, for all 40 SNe, in time for the Jan. 8, 1998 AAS meeting. 

For confirmation, I asked  my colleagues to check these  results,  in case there could have been  a mistake.  However, all the 20 entries in the so called ``Gerson-table'' for each of the 38 SNe were confirmed as correct. Later Greg Aldering added more columns to this  table, giving the correlated errors, and Richard Ellis added a column with the best estimate of the nature of each host galaxy.

\begin{figure}[!thbp]
\resizebox{0.8\textwidth}{!}{\includegraphics{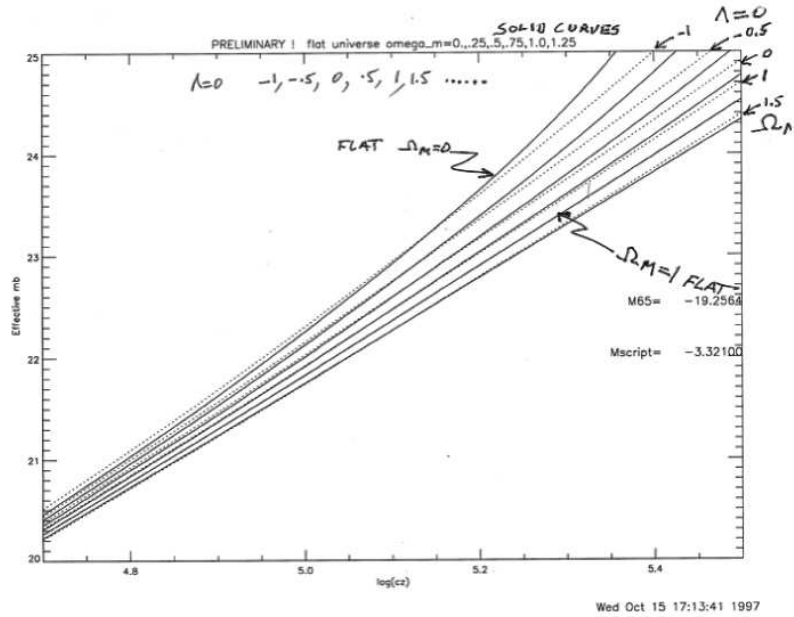}}
\caption{The Hubble curves for both a flat universe (solid curves) and an  $\Omega_\Lambda = 0 $ universe  (dotted). }   
\label{fig:5}
\end{figure}

\begin{figure}[!thbp]
\resizebox{0.45\textwidth}{!}{\includegraphics{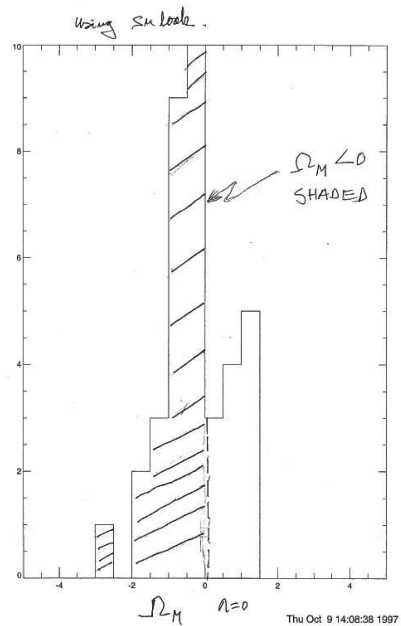}}
\caption{ The $\Omega_m$ distribution for a $\Omega_\Lambda = 0 $ universe. Note that 25 of the 38 SNe have a negative  $\Omega_m$ and hence an un-physical value. This shows clearly and independently of the flat universe condition that  $\Omega_\Lambda$ must be greater than zero. }   
\label{fig:6}
\end{figure}

Although as mentioned above,  Goobar and Perlmutter (1995)\cite{goobar95a} had shown that any SN measurements indicating low $\Omega_m$ for a flat universe will also indicate negative $\Omega_m$ for a $\Omega_\Lambda=0$ universe, it was striking to see this directly in the data.  Thus, in Fig.~\ref{fig:5}. the Hubble curves for both a flat universe and an $\Omega_\Lambda = 0$ universe are shown superimposed. For an   $\Omega_\Lambda = 0$    universe  we  obtain a corresponding  peak on the $\Omega_m$   histogram. However now it occurs for negative  $\Omega_m$ values. In Fig.~\ref{fig:6}. is shown the $\Omega_m$ histogram, this time  calculated with Saul's SNLOOK program, for this case. As can be noted, 25 out of the 38  SNe give a negative and hence un-physical value for $\Omega_m$. This clearly demonstrates that $\Omega_\Lambda$ cannot be zero but must  be greater than zero. Figures~\ref{fig:5} and  \ref{fig:6} are from my notebook from October  1997 and were shown informally to group members. This result was also presented in my seminar at Santa Barbara in December, 1997.

One can also turn  this argument around and calculate the location of the peak in an $\Omega_m$ histogram for a series of different  increasing $\Omega_\Lambda$ values and find for which $\Omega_\Lambda$ value the peak begins to lie at positive (hence physical) values. This approaches, and goes beyond,   $\Omega_m$, $\Omega_\Lambda$ values consistent with the above values for a flat universe. This method  is equivalent to traveling along the central line for the confidence level ``ellipses'', starting at negative $\Omega_m$ values, shown in Fig.~\ref{fig:10}.

We thus have two ways to visualize with histograms the evidence for a positive value for $\Omega_\Lambda$. The first histogram essentially looks at where the diagonal confidence region (in, e.g., Fig.~\ref{fig:10}) crosses the flat universe line, and the second looks at where it crosses the $\Omega_\Lambda = 0$ axis. Both indicated that $\Omega_\Lambda$  must be positive (since the $\Omega_\Lambda = 0$ histogram gives unphysical results). And, of course, when we calculated $q_0$ the supernova data clearly indicated acceleration. 

In Table 1 are shown the SN discoveries over the years 1989 to 1997. Saul's SNe on demand  method really came into its own when we obtained time on the Cerro/Tololo 4m telescope combined with the ``big throughput''  camera  (BTC) of Bernstein and Tyson. Table 1 shows that  over a nine-year period we discovered  42 SNe with $z  >  0.2$.

\begin{table}
\begin{tabular}{lccc}
\hline
Epoch   & Number & Total & Telescope             \\
\hline
1989-91 &  0     &    0  & Anglo Australian 3.9m \\
1992    &  1     &    1  & La Palma 2.5m         \\
1993    &  6     &    7  & Kit Peak 4m           \\
1995    &  9     &   16  & Cerro Tololo 4m       \\
1996    &  7     &   23  & Cerro Tololo 4m       \\
1997    & 19     &   42  & Cerro Tololo 4m       \\
\hline
\end{tabular}
\caption{SCP SNe~Ia Discoveries with $z > 0.2$ }
\label{tab:1}
\end{table}

\subsection{Oct. 1997. Two papers by the SCP and Hi-Z Team as well as  a joint paper submitted with  hints, from very small samples of SNe, for a non-zero $\Omega_\Lambda$}

\begin{figure}[!thbp]
\resizebox{0.75\textwidth}{!}{\includegraphics{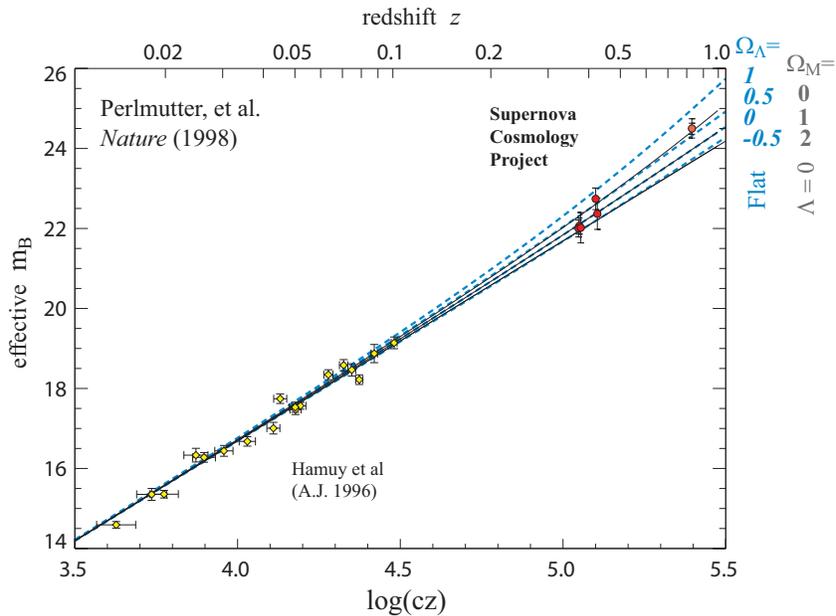}}
\caption{Hubble diagram from the Nature paper\cite{perlmutter98a} showing the effect of the very well-measured SN at  $z=0.83$ on the  determination of cosmological parameters. }
\label{fig:7}
\end{figure}

The very first analysis of multiple high-redshift SNe indicated a high $\Omega_m$ value, albeit with large statistical error bars: using the 5 of these first 7 SNe with lightcurve timescales that were calibrated at low redshift yielded $\Omega_m = 0.94 ^{+ 0.34} _{- 0.28}$ for a flat universe \cite{perlmutter95b,ruiz-lapente97a,perlmutter97a}.  The following year, we added one very-well-observed type Ia supernovae that we discovered and followed with the Hubble Space Telescope (HST), SN 1997ap, that was at that time again the most
distant that had ever been observed, $z = 0.83$, with the most leverage for this measurement. The addition of this single SN    allowed us to recalculate 
$\Omega_m$ for 5+1 supernovae, giving $\Omega_m = 0.6 \pm 0.2$ for a flat universe -- for the first time in a publication a direct indication of a positive cosmological constant, albeit only at the two sigma level. Fig.~\ref{fig:7} shows the Hubble diagram from this paper, where the new well-measured highest redshift SNe indicates a positive cosmological constant with even a stronger significance.  We submitted this paper to {\it Nature} on Oct. 6, 1997 \cite{perlmutter98a} and it appeared in the January 1, 1998 issue.

Garnavich et al. \cite{garnavich98a} submitted a paper by the Hi-Z Team on Oct. 13, 1997, based on three confirmed high redshift SNe measured with the HST, and one measured from the ground. They concluded that  matter alone is insufficient to produce a flat universe. Specifically for $\Omega_m  +  \Omega_\Lambda = 1 $ , $\Omega_m$  is less than 1 with more than 95\%  confidence, and their  best estimate of $\Omega_m$ is $-0.1 \pm 0.5$ if $\Omega_\Lambda = 0$.

A joint paper by some members of the two groups Riess et al. \cite{riess98b} submitted Nov. 21, 1997 and accepted Apr. 17, 1998, showed that the ``snapshot''  method works. That involves taking a single photometry point and a single spectrum on the same night and deducing the magnitude at maximum light from this limited information. On the basis of four SNe handled in this fashion, as well as the SNe from the above two papers,  they deduced that for a flat universe $\Omega_m = 0.19 ^{+ 0.32} _{- 0.19}$ and $\Omega_m = -0.31 ^{+ 0.62} _{- 0.36}$ for an  $\Omega_\Lambda = 0$ universe.

While these  SCP and Hi-Z  papers, published in 1998,  gave hints of a non-zero $\Omega_\Lambda$ they relied on a very small number of SNe. The hints from the snapshot paper similarly relied on four  SNe using this less precise method.

All this work was going on more or less contemporaneously with my study of the 38 SNe. My colleague Peter Nugent, who did some of the fits of  these confidence level distributions, felt that these 3 ``hints'' were a more convincing and possibly earlier evidence for a non-zero $\Omega_\Lambda$ than the peak in the $\Omega_m$  histogram. My feeling is that we needed the larger statistics since we had amply demonstrated that with a very small number of SNe (7 SNe) one can be way off!

\section{Oct. 23 to Dec. 14,  1997: Three Colloquia and a Seminar}

{\it The first four public presentations of the SCP evidence for $\Omega_\Lambda > 0$ and acceleration, based on the peak in the $\Omega_m$  histogram for 40 SNe as well as the unphysical results obtained  when $\Omega_\Lambda$ is assumed to be zero.}
\vspace{0.2in}

From October to November, we re-measured and re-fitted all of the 38 SNe and added two more, giving 40 (while three were identified as outliers). We also changed from the Burstein and Heiles \cite{burstein82a} version of dust extinction in our galaxy to the more recent Schlegel, Finkbeiner and Davis \cite{schlegel98a} version. I then revised the tables  to reflect all these changes. Both Saul and I plotted the Hubble plots, based on the  revised tables, and used SNLOOK to obtain the latest version of the histogram.  

\begin{figure}[!thbp]
\resizebox{0.5\textwidth}{!}{\includegraphics{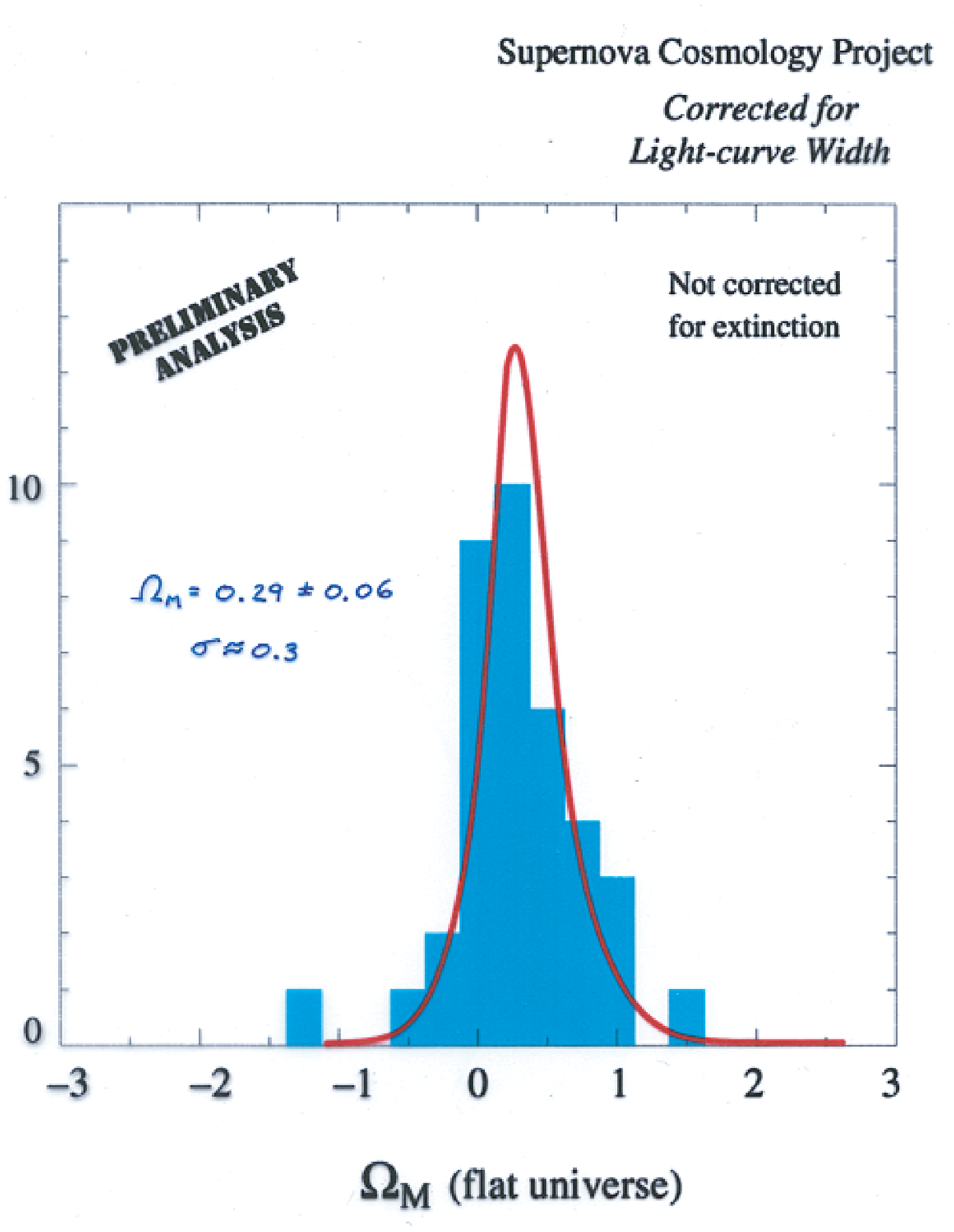}} 
\resizebox{0.5\textwidth}{!}{\includegraphics{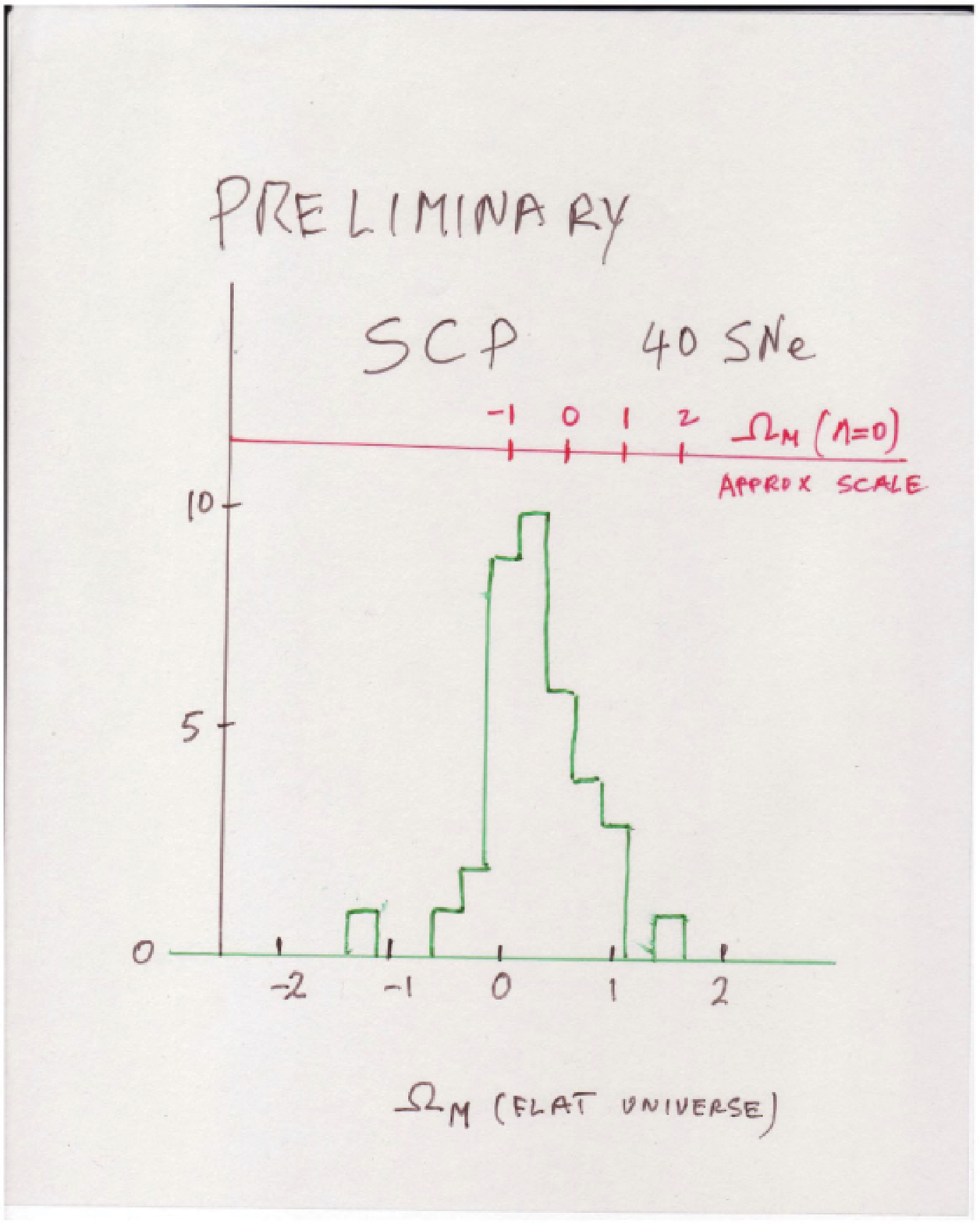}} 
\caption{ (\emph{left}) The  figure of the $\Omega_m$ distribution shown at Colloquia at UC San Diego in October 1997 and at  UC Berkeley and Santa Cruz in early December 1997  by Saul Perlmutter. The x-axis  gives $\Omega_m$ for a flat universe. All the SNe given in Fig.~\ref{fig:4}. have been re-fitted. The Gaussian fit to this preliminary data gave  $\Omega_m = 0.29 \pm 0.06$ with a $\sigma = 0.3$.  This analysis accounts for
dust  with the more correct statistical approach discussed in the text above concerning Fig. 4, rather than correcting extinction SN by SN.
(\emph{right}) A reproduction of the actual figure of the $\Omega_m$ distribution shown at the ITP Santa Barbara on Dec. 14, 1997 by Gerson Goldhaber. The lower scale gives $\Omega_m$ for a flat universe. The upper scale gives  $\Omega_m$ for an  $\Omega_\Lambda = 0 $ universe. Note that this assumption gives negative (and thus unphysical) values for $\Omega_m$. This demonstrates that  $\Omega_\Lambda $ must be greater than zero by a method which does not require the flat universe condition.}   
\label{fig:8and9}
\end{figure}

The first public presentation of our results was a colloquium by Saul on October 23, 1997 at the Physics Department, UC San Diego.  This was  followed by Saul's colloquia at the Physics Department UC Berkeley on Dec. 1, 1997 and at the Physics Department  UC Santa Cruz on Dec. 11, 1997  (See Fig.~\ref{fig:8and9}).

Next, I gave a seminar  at the Institute for Theoretical Physics (ITP) at UC Santa Barbara on  Dec. 14, 1997 (See Fig.~\ref{fig:8and9}).  In all four of these talks, as stated above, the conventional description of the confidence level  distribution on the $\Omega_m $ , $\Omega_\Lambda$  plane was not yet available for the 40 SNe. So in all four talks the  $\Omega_m $ histograms, re-evaluated  with more careful re-measurements, were shown. We showed that the result was robust against the likely sources of systematic error, but we still considered our result  as preliminary as we had not yet completely explored every possible effect. 

It was not easy to convince the astrophysics community of this result, as it was contrary to all ingrained beliefs.\footnote{Examples of  exceptions to the $\Omega_\Lambda = 0$ assumption are: Steve Weinberg \cite{weinberg89a} made an argument, on the basis of the anthropic principle, that $\Omega_\Lambda$ could be comparable to the currently observed value. Krauss and Turner \cite{krauss95a} argued primarily on the basis of the age of the universe and other then available cosmological measurements that  $\Omega_\Lambda$ should lie between 0.6 and 0.7. Ostriker and Steinhardt \cite{ostriker95a} derived a concordance model based on  all the then available cosmological data and concluded  $\Omega_\Lambda = 0.65 \pm 0.1$.} Acceleration rather than deceleration as expected!

The question has been posed: who remembers the details of these talks? I know that Kim Griest remembers Saul's talk at UC San Diego and was enthusiastic about the result. I also know Joel Primack and Michael Riordan remember Saul's talk at UC Santa Cruz. As former and current particle physicists they understood the significance of a  peak on a histogram. Joel Primack was particularly enthusiastic and stressed the importance of  this discovery after Saul's colloquium. 
Dave Branch remembers my talk and subsequent discussions and understood its significance. David Gross asked me after my talk how I could come to  such a momentous conclusion  on the basis of just 40 SNe. 

As is usual,  Saul's Colloquium at UC Berkeley was video taped (DVD available). Saul first talked about the ``hint'' from our 5+1 SNe fit I mentioned above. Then based on a gaussian fit to the observed peak on the histogram, Saul quoted  $\Omega_m  = 0.29 \pm 0.06 $ for a flat universe and thus $\Omega_\Lambda = 0.71$. In my presentation at the ITP I did not try for an accurate value of  $\Omega_m $, but rather showed Fig.~\ref{fig:8and9} and indicated that $\Omega_m  = 0.3$ agrees with our data for a flat universe. In addition I demonstrated what happens when we assume $\Omega_\Lambda  = 0$. As seen from the upper scale in Fig.~\ref{fig:8and9} this assumption gives negative and hence unphysical values for  $\Omega_m $.
As stated earlier,  given the slope of  the confidence  region for any  supernova data,  such values for $\Omega_m$ in the  two cases  considered  demonstrate that  $\Omega_\Lambda$  must be greater  than zero, and does not depend on  assuming a flat  universe.

In his talk Saul  went on to point out that while on the one hand 0.7 for $\Omega_\Lambda $ corresponded to  a large fraction of the universe but on the other hand it was a very small value, by a factor  $10^{122}$, compared to the value expected from virtual particle vacuum fluctuations. 
  
It is remakable that in Saul's  talk on Dec. 1, 1997 we  had already  presented, in public,  evidence for the whole story of what was later called Dark Energy by Michael Turner.  At DM08 I presented a four-minute video excerpt of Saul's talk, showing him making these points based on the histogram. (This video is given in the DM08 conference proceedings \cite{perlmutter97c}.) 
  
Not much has changed qualitatively in the 10 years since then. With many more SNe studies \cite{kowalski08a} as well as concordance with CMB and evidence from cluster studies \cite{bahcall00a} and later Baryon Acoustic Oscillation (BAO) \cite{eisenstein05a} studies, we now know  $\Omega_m $ to  more significant figures, but qualitatively there is no change from the value we quoted in 1997. We have learned that the equation of state parameter $w$ is very close to -1 and have some limits on the variation of $w$ with z. But after all that, just as 10 years ago,  we still do not understand the nature of Dark Energy \cite{rubin08a}.

\section{Jan. 8, 1998: First Public Presentation of the SCP results at an  American Astronomical Society meeting}

{\it The fifth  public presentation of the SCP evidence for  $\Omega_\Lambda > 0$ and acceleration, based on the fit to the confidence level distribution on the $\Omega_m $ , $\Omega_\Lambda$  plane for 40 SNe.}
\vspace{0.2in}
 
From October to late in December 1997, the SCP group at Berkeley re-measured and re-checked and completed all error and correlated error calculations on  what had by then become 40 SNe. This effort was led by  Saul, Greg Aldering,  Peter Nugent, and Rob Knop.  Each group member as well as the group members in Baltimore, Stockholm,  Paris, England, Australia, Spain and Chile convinced himself/herself by direct study of the data that indeed $\Omega_\Lambda > 0 $ and that contrary to  preconceived notions we were dealing with acceleration of the universe, namely a negative deceleration coefficient. The collaboration was prepared to make a major presentation.
 
\begin{figure}[!thbp]
\resizebox{0.5\textwidth}{!}{\includegraphics{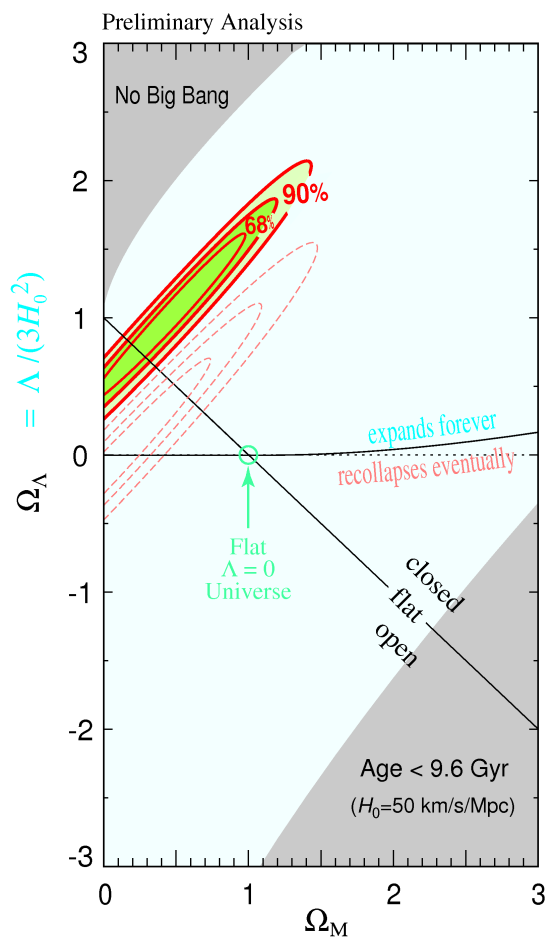}} 
\caption{ The $\Omega_\Lambda$ vs $\Omega_m$ best fit confidence level distribution shown at the Washington meeting of the AAS on Jan, 8 1998   by Saul Perlmutter. The dashed curves represent a very generous estimate of what possible systematic errors could do. }   
\label{fig:10}
\end{figure}

The following month, on Jan. 8, 1998  \cite{perlmutter98b,perlmutter97b}, Saul Perlmutter presented our data at the AAS meeting in Washington DC. By this time we had  the  much more elaborate analysis, based on  detailed error calculations,  by the entire SCP group. We had 40 fully analyzed SNe and the confidence level calculation was already available on the $\Omega_m $, $\Omega_\Lambda$  plane (Fig.~\ref{fig:10}). 
Here it should be noted that the confidence level distribution does not require the  flat universe condition to give $\Omega_\Lambda$ greater than zero. This was the first presentation at an official conference of evidence for $\Omega_\Lambda > 0$, and hence from finding  $q_0 < 0 $,  the acceleration of the expansion of the universe.  To take possible systematic uncertainties  into account, we allowed for a very generous systematic error and they were shown for the case where all systematics conspired in the same direction. This is shown by the dotted confidence level curves in Fig.~\ref{fig:10}. This figure was reproduced by James Glanz in his report on our presentation in Science \cite{glanz98a}. Furthermore, using the same  $\Omega_m$ histogram  for a flat universe as in Fig.~\ref{fig:8and9},  Saul also showed the studies of the effects of systematics: Malmquist bias, stretch dependence, and color/dust dependence.

\section{Feb. 18, 1998: Public presentations at the Dark Matter 1998 (DM98) Conference.} 

{\it The sixth and seventh   public presentations of the SCP evidence for $\Omega_\Lambda > 0$ and acceleration based on 42 high redshift SNe were given. The first public presentation  by the Hi-Z Team of  evidence for $\Omega_\Lambda > 0$ and acceleration based on 14 independent high redshift  SNe was also given. Both teams quoted results based on  fits to the confidence level distribution on the $\Omega_m $ , $\Omega_\Lambda$  plane. }
\vspace{0.2in}
  
The following month, on Feb. 18,  1998,  both I and Saul Perlmutter gave talks,  in that order \cite{goldhaber98a}, showing our results at the meeting organized by Dave Cline just 10 years ago (Dark Matter 1998) at Marina del Rey. By this time our number of fully analyzed SNe had grown to 42. ( See Table 1.) 

I spent part of my talk on some work I  had done on time dilation in SNe explosions, proving that the redshift is due to the expansion of the universe, rather then to the ``tired light'' hypothesis \cite{perlmutter95b,ruiz-lapente97a,goldhaber01a}.  I then  showed the best fit confidence level distribution on the $\Omega_m$, $\Omega_\Lambda$  plane, which for a flat universe implied $\Omega_m   = 0.28^{+ 0.09}_{-0.08}  $ (statistical)  $^{+ 0.05}_{- 0.04}$ (systematic) and hence, from evaluating $q_0$, acceleration of the universe at the present epoch. 

In his talk Saul Perlmutter gave more  details on our results  and on the error calculations, and stressed  the studies of the systematic uncertainties.

Following our talks, Alexei Filippenko presented the results \cite{filippenko98a} of the Hi-Z Team, who  claimed they had established the  acceleration of the universe on the basis of a confidence level  analysis of 16 high redshift SNe. These consisted of 10 well measured  SNe, plus 4 from the ``snapshot'' method \cite{riess98b}  plus 2 which came from our set of 42 \cite{perlmutter99a}. 

Later the refereed publications of the two groups capped these momentous results. The Hi-Z Team paper was submitted on Mar. 13, 1998 \cite{riess98a} and the SCP paper was submitted on Sept. 8, 1998 \cite{perlmutter99a}.

\section{How does an $\Omega_m$ histogram look today?}

\begin{figure}[!ht]
\resizebox{0.5\textwidth}{!}{\includegraphics{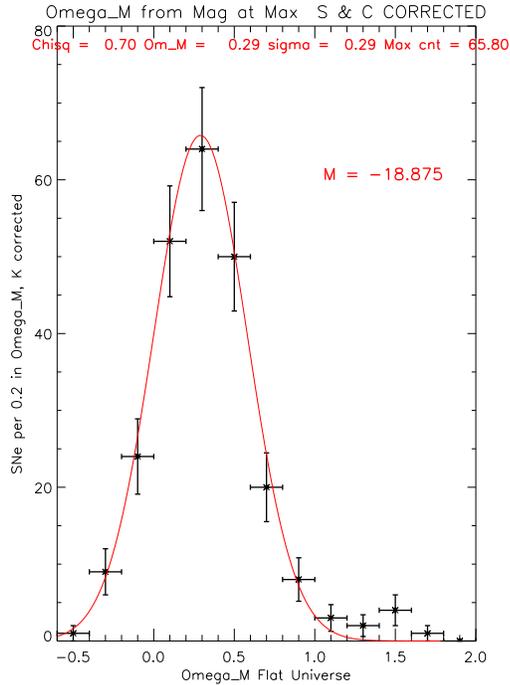}}
\caption{ The $\Omega_m$ distribution for a sample of 257 SNe with $z > 0.2$ out of a total of $ 307 $ which were collected and re-analyzed in our group  by Marek Kowalski, David Rubin et al. [29] in connection with a current (2008) SCP paper giving a Union of all published SNe. The SN magnitudes at maximum light were K-corrected as well as stretch and color corrected. Here as above   $\Omega_\Lambda$ is given by $1- \Omega_m$.}
\label{fig:11}
\end{figure}

To see how  the method, patterned on  particle physics and  used to indicate  acceleration on Sept. 24, 1997, works  with larger statistics I have analyzed (this time with a computer program)  a sample of 257 SNe with $ z > 0.2$ [28]  in the same fashion as in Figs.~\ref{fig:3} and \ref{fig:4}.  Fig.~\ref{fig:11} shows the resulting  distribution in $\Omega_m$. The  enormous  peak is clearly observed, which completely confirms the validity of the peak for the 38 SNe in Fig.~\ref{fig:4}.

\section{Feb. 20, 2008: So what is the ``Fate of the Universe'' ?}

By now, 10 years later, there are many experiments which have confirmed  Dark Energy. We can however still ask: what is the "Fate of the Universe" ?

The question is now in the form: is $\Omega_\Lambda$ constant or does it vary with $ z$? Or alternatively: is $w =  -1$, namely are we dealing with  Einstein's cosmological constant ?         Or can one measure  deviations from $-1$ where $w$ is  the equation of state parameter?

We hope the Joint Dark Energy Mission (JDEM)  will answer this question before the next decade is up at DM2018.

\section{Summary of the steps in the SCP discovery of evidence for Dark Energy}
 
\begin{itemize}
\item  1988 first pilot search for high-redshift SNe at Palomar organized by Carl Pennypacker and Saul Perlmutter.
\item  1988 proposal to NSF center at UCB for ``Fate of Universe" study. 
\item 1992 discovery of SN1992bi, z = 0.486. 
\item 1995 first 7 SNe yielded large  $\Omega_m$    value. 
\item   Sept. 24,  1997  $\Omega_m$   = 0.2 to 0.3 for flat universe from peak in histogram, SCP group meeting based on 38 SNe.
\item   Oct. 1997 hints for non zero  $\Omega_\Lambda$  from SCP and Hi-Z Team.
\item  Oct. 23 to Dec. 14,  1997 public talks by Saul Perlmutter and Gerson Goldhaber showing evidence that for a flat Universe  $\Omega_m$   =  0.3 and hence  $\Omega_\Lambda$  = 0.7 based on the peak in the $\Omega_m$ histogram, Figs.~\ref{fig:8and9}, and by then on 40 SNe.  
\item Jan. 8,  1998 SCP presentation by Saul Perlmutter at the AAS meeting in Washington DC showing the $\Omega_m$ ,    $\Omega_\Lambda$  plane with best fit confidence level distributions yielding flat universe values of   $\Omega_m  = 0.3$ and  $\Omega_\Lambda  = 0.7$ . 
\item  Feb. 18,  1998 Both groups talk at Marina del Rey DM98. Gerson and Saul show evidence for  a non-zero $\Omega_\Lambda$  and acceleration based on 42 SNe, followed by  Filippenko who showed evidence for a non-zero $\Omega_\Lambda$  and acceleration based on 10 + 4 + 2  SNe for the Hi-Z Team.
\end{itemize}

\section{Acknowledgements}

I wish to thank Kyle Barbary, Alex Gude, David Rubin, Tony Spadafora and Nao Suzuki for  help with this manuscript.
 
This work has been supported in part by the Director, Office of Science, Office of High Energy Physics, of the U.S. Department of Energy under Contract No. DE-AC02-05CH11231.

\bibliographystyle{aipprocl}
\bibliography{refs}

\end{document}